\documentclass[%
 aip,
amsmath,amssymb,
 reprint,%
]{revtex4-1}

\usepackage{graphicx}
\usepackage{dcolumn}
\usepackage{bm}
\usepackage{amsmath}
\usepackage[utf8]{inputenc}
\usepackage[T1]{fontenc}
\usepackage{mathptmx}
\usepackage{etoolbox}
\usepackage[normalem]{ulem}
\usepackage{multirow}
\usepackage{comment}

\usepackage[usenames,dvipsnames]{color} 
\usepackage[colorlinks=true,linkcolor=blue,citecolor=blue]{hyperref}%

\makeatletter
\def\@email#1#2{%
 \endgroup
 \patchcmd{\titleblock@produce}
  {\frontmatter@RRAPformat}
  {\frontmatter@RRAPformat{\produce@RRAP{*#1\href{mailto:#2}{#2}}}\frontmatter@RRAPformat}
  {}{}
}%
\makeatother
\begin{document}

\preprint{AIP/123-QED}

\title[]{Near- to mid-IR spectral purity transfer with a tunable frequency comb: methanol frequency metrology over a 1.4~GHz span}
\author{D. B. A. Tran}
\altaffiliation{Current address: Time and Frequency Department, National Physical Laboratory, Teddington, United Kingdom}
 \affiliation{Laboratoire de Physique des Lasers, Université Sorbonne Paris Nord, CNRS, Villetaneuse, France}
 \affiliation{Faculty of Physics, Ho Chi Minh City University of Education, Ho Chi Minh City, Vietnam}
 
\author{O. Lopez}%
 \affiliation{Laboratoire de Physique des Lasers, Université Sorbonne Paris Nord, CNRS, Villetaneuse, France}
 \author{M. Manceau}%
 \affiliation{Laboratoire de Physique des Lasers, Université Sorbonne Paris Nord, CNRS, Villetaneuse, France}
  \author{A. Goncharov}%
\altaffiliation{Permanent address: Institute of Laser Physics, Siberian Branch of the Russian Academy of Sciences, Novosibirsk, Russia}
 \affiliation{Laboratoire de Physique des Lasers, Université Sorbonne Paris Nord, CNRS, Villetaneuse, France}
 
\author{M. Abgrall}
\affiliation{LNE-SYRTE, Observatoire de Paris, Université PSL, CNRS, Sorbonne Université, Paris, France}
\author{H. Alvarez-Martinez}
\affiliation{LNE-SYRTE, Observatoire de Paris, Université PSL, CNRS, Sorbonne Université, Paris, France}
\affiliation{Real Instituto y Observatorio de la Armada, San Fernando, Spain}
\author{R. Le Targat}
\affiliation{LNE-SYRTE, Observatoire de Paris, Université PSL, CNRS, Sorbonne Université, Paris, France}
\author{E. Cantin}
 \altaffiliation{Current address: Laboratoire de Physique des Lasers, Université Sorbonne Paris Nord, CNRS, Villetaneuse, France}
\affiliation{LNE-SYRTE, Observatoire de Paris, Université PSL, CNRS, Sorbonne Université, Paris, France}
\author{P.-E. Pottie}
\affiliation{LNE-SYRTE, Observatoire de Paris, Université PSL, CNRS, Sorbonne Université, Paris, France}
\author{A. Amy-Klein}
 \affiliation{Laboratoire de Physique des Lasers, Université Sorbonne Paris Nord, CNRS, Villetaneuse, France}
\author{B. Darquié}
\email{benoit.darquie@univ-paris13.fr}
 \affiliation{Laboratoire de Physique des Lasers, Université Sorbonne Paris Nord, CNRS, Villetaneuse, France}

\date{\today}

\begin{abstract}
We report the upgrade and operation of a frequency-comb-assisted high-resolution  mid-infrared molecular spectrometer allowing us to combine high spectral purity, SI-traceability, wide tunability and high sensitivity. An optical frequency comb is used to transfer the spectral purity of a SI-traceable 1.54 µm metrology-grade frequency reference to a 10.3~µm quantum cascade laser (QCL). The near-infrared reference is operated at the French time/frequency metrology institute, calibrated there to primary frequency standards, and transferred to Laboratoire de Physique des Lasers \textit{via} the REFIMEVE fiber network. The QCL exhibits a line width of $\delta\nu\sim0.1$~Hz, a sub-$10^{-15}$ relative frequency stability from 0.1 to 10 s and its frequency is traceable to the SI with a total relative uncertainty better than $4\times 10^{-14}$ after 1-s averaging time. We have developed the instrumentation allowing comb modes to be continuously tuned over 9 GHz resulting in a QCL of record spectral purity uninterruptedly tunable at the precision of the reference over an unprecedented span of  $\Delta\nu=1.4$~GHz. We have used our apparatus to conduct sub-Doppler spectroscopy of methanol in a multi-pass cell, demonstrating state-of-art frequency uncertainties down to the few kilohertz level ($\sim10^{-10}$ in relative value). We have observed weak intensity resonances unreported so far, resolved subtle doublets never seen before and brought to light discrepancies with the HITRAN database. This demonstrates the potential of our apparatus for probing subtle internal molecular processes, building accurate spectroscopic models of polyatomic molecules of atmospheric or astrophysical interest, and carrying out precise spectroscopic tests of fundamental physics.
\end{abstract}

\maketitle

\section{\label{sec_1}Introduction}
High precision molecular spectroscopy offers exciting perspectives ranging from fundamental physics \cite{hinds1997testing,roussy2023improved,fiechter2022toward,pastor2015testing,patra2020proton,diouf2019lamb,barontini2022measuring,muller2021study,segal2018studying,mudiayi2021linear} and metrology \cite{benabid2005compact,castrillo2019optical,mejri2015measuring} to astrochemistry \cite{herbst2009complex}, remote sensing and Earth sciences \cite{galli2016spectroscopic,harrison2011spectroscopic,guinet2010laser}. Experiments in these domains are often based on frequency measurements of molecular vibrations in the mid-infrared (mid-IR) molecular fingerprint region, therefore generating the need for mid-IR laser sources that are spectrally pure, accurate and widely-tunable.

Here we report on the upgrade and operation of a quantum cascade laser (QCL) based spectrometer that provides a unique combination of sensitivity, frequency resolution and tunability of any mid-IR spectrometer to date. An Erbium-doped fiber mode-locked optical frequency comb is used to transfer the spectral purity of a metrology-grade ultrastable frequency reference traceable to the International System of Units (SI) from the near-IR to the mid-IR. In addition, we have developed the instrumentation allowing comb modes to be continuously tuned over 9 GHz. This results in a QCL of spectral purity at the state-of-the-art that can be continuously tuned at the precision of the frequency reference over an unprecedented frequency span of 1.4 GHz. These developments constitute enabling technologies for driving the next generation of ultra-high resolution molecular spectroscopy apparatus in the molecular fingerprint region. High spectral purity is key to reaching the resolutions required for resolving the subtlest structures, clusters of blended lines which constitute unique probes of fundamental processes in molecules. Together with SI-traceability and high sensitivity, it allows systematic effects to be unraveled in order to minimize line center uncertainties. Adding finally wide continuous tuning capabilities to the toolbox of present and future mid-IR photonics is important to facilitate the discovery/resolution of new unreported lines, in particular weak transitions, as demonstrated here, providing in-depth insight into the internal molecular dynamics. 

Distributed feedback (DFB) quantum cascade lasers (QCLs) are available over wide ranges of the mid-IR (from 3 to 25 µm), can be tuned over several  hundreds of gigahertz and have continuous-wave (cw) power levels in the milliwatt to watt range near room-temperature. They however show substantial frequency noise~\cite{myers2002free,bartalini2010observing,tombez2011frequency,bartalini2011measuring,tombez2012linewidth,mills2012coherent,galli2013comb,sow2014widely,argence2015quantum,cappelli2015intrinsic,chomet2023highly,manceau_demonstration_2023}. For the most precise measurements, high-spectral purity and traceability to a frequency standard are both required. This can be achieved by phase-locking to: (i) the mid-IR secondary frequency standard \cite{sow2014widely} (a CO$_{2}$ laser stabilized to a saturated absorption line of OsO$_{4}$ \cite{acef1997metrological}), resulting in a $\delta\nu=10$~Hz line width, 1 Hz stability at 1 s and accuracy of a few tens of hertz; (ii) a near-IR metrology-grade frequency reference traceable to primary frequency standards, resulting in ultimate sub-Hz stabilities and accuracies, and linewidth narrowing down to the $\delta\nu=0.1$~Hz level \cite{santagata2019high,argence2015quantum} (see also related similar works, which however do not reach such high levels of spectral purity \cite{insero2017measuring,sinhal2022frequency}). Efforts in combining sub-100~Hz metrology-grade spectral performance and wide continuous tunability are scarce. We have previously demonstrated respectively $\Delta\nu=10$~GHz and $\Delta\nu=0.4$~GHz continuous tunability for: (i) the 10-Hz narrow 10.6 µm QCL phase-locked to the mid-IR secondary frequency standard ($\sim10^{-12}$ absolute frequency uncertainty) \cite{sow2014widely} and (ii) the 0.1-Hz narrow 10.3 µm QCL ultimately calibrated to primary frequency standards ($\sim10^{-14}$ absolute frequency uncertainty) \cite{santagata2019high}. Another interesting apparatus is the CO$_{2}$-laser/microwave-sideband spectrometer demonstrated in \cite{sun2004dual} allowing 11.8 GHz wide windows around any CO$_{2}$ laser emission line to be covered in a single sweep. This spectrometer shows similarities with our sub-100~Hz metrology-grade equipment \cite{sow2014widely,lemarchand2013revised}, but uses free-running CO$_{2}$ lasers and allows broadband scans at Doppler-limited resolutions only. Note that QCLs can also be phase-locked to a frequency comb controlled with a radiofrequency (RF) reference. This enables higher tuning ranges, but at the expense of a barely reduced frequency noise (resulting in $\sim$1 kHz line width at best) and a frequency uncertainty limited to around $10^{-13}$ at best by the RF reference~\cite{bartalini2007frequency,gambetta2010frequency,mills2012coherent,borri2012direct,gatti2013frequency,galli2013comb,gambetta2015direct,hansen2015quantum,gambetta2017absolute,lamperti2018absolute,lamperti_optical_2020}. Mid-IR frequency combs, based~\cite{lepere2022mid,agner2022high,komagata_coherent_2023,parriaux_coherence_2023,gabbrielli_timefrequency-domain_2023} or not~\cite{timmers_molecular_2018,muraviev_massively_2018,krzempek_stabilized_2019,ruehl_widely-tunable_2012,chomet2023highly,laumer2023sub,gabbrielli_timefrequency-domain_2023} on QCL technologies, are emerging as flexible sources of broadband coherent radiation. They can address a very wide spectral range, via dual-comb~\cite{karlovets_dual-comb_2020} or Fourier transform~\cite{hjalten_optical_2021,yang2023self} spectroscopy methods \cite{picque2019frequency}, but sub-100~Hz metrology-grade performance (line width, stability, accuracy) have yet to be demonstrated for mid-IR combs.

In the following, we describe our high-resolution molecular spectrometer. It exploits a frequency comb to transfer the spectral purity of a 1.54 µm remote SI-traceable optical frequency reference signal to a 10.3 µm QCL. The near-IR reference is operated at the French time/frequency metrology institute (LNE-SYRTE), calibrated there to primary frequency standards, and transferred to Laboratoire de Physique des Lasers (LPL) via a long-haul optical fiber link \cite{argence2015quantum}. Compared to our previous work\cite{santagata2019high}, the QCL's tuneability has been extended by a factor of more than three. We use our SI-traceable sub-Hz QCL to conduct sub-Doppler saturation spectroscopy of methanol in a multi-pass cell over an unprecedented continuous frequency span of $\Delta\nu=1.4$~GHz. This corresponds to a record continuous relative tuning range of $\Delta\nu/\delta\nu\sim10^{10}$, with $\Delta\nu/\delta\nu$ equal to the ratio of the spectral coverage to the instrumental resolution given by the laser line width. 

Methanol is found in a wide variety of astronomical sources \cite{batrla1987detection} and is the second most abundant organic molecule in the Earth’s atmosphere \cite{singh2001evidence} . It is an excellent probe of the physical conditions, the chemistry and the history of these environments. Reliable laboratory spectroscopic data of methanol are thus much needed for interpreting astrophysical and planetary spectra, for air quality monitoring and atmospheric concentration retrieval. Although the simplest of the organic alcohols, it is a slightly asymmetric-top with a hindered internal rotor (torsion) leading to a triple-well internal tunnel dynamics and therefore to a rich rotation–torsion–vibration energy structure. Methanol is thus also an important molecule for fundamental spectroscopy\cite{lankhaar2016hyperfine}, metrological applications and frequency calibration \cite{santagata2019high,sun2000sub}, the realization of optically pumped far-IR gas lasers \cite{jiu2010pulsed}, or for probing the limits of the standard model, its internal tunnel dynamics making it one of the most sensitive molecules for a search of a varying proton-to-electron mass ratio \cite{muller2021study}. Here, we measure the resonance frequencies of fourteen rovibrational transitions, including very weak lines, some observed for the first time. We demonstrate state-of-the-art frequency uncertainties two to four orders of magnitude improved over previous measurements, at the few kilohertz level ($\sim10^{-10}$ in relative value) for the most intense lines and bring to light inaccuracies and gaps in the HITRAN database. We resolve subtle weak intensity doublets never observed before induced by a combination of the asymmetry and the tunneling dynamics in methanol, demonstrating both the high detection-sensitivity and high resolution of our widely tunable spectrometer and its potential for unraveling subtle internal molecular processes.

\section{Experimental setup}

FIG. \ref{fig_1} illustrates our widely tunable SI-traceable frequency-comb-stabilized-QCL-based high-resolution mid-IR spectrometer. It combines a widely tunable optical local oscillator (OLO) locked to a remote ultrastable frequency reference of LNE-SYRTE, a 10.3 µm cw DFB QCL (Alpes Lasers, LLH package) stabilized to the OLO using an optical frequency comb (OFC) and sum-frequency generation, and a multi-pass absorption cell for carrying out saturation spectroscopy. The reader is referred to Ref. \cite{santagata2019high} for a detailed description of most of this setup. Note that most of the instrumentation described in this section, in particular all the locking electronics is home-made.

\subsection{\label{sec2_1}Widely tunable, SI-traceable and ultrastable optical local oscillator}

\begin{figure*}
\includegraphics[width=\textwidth]{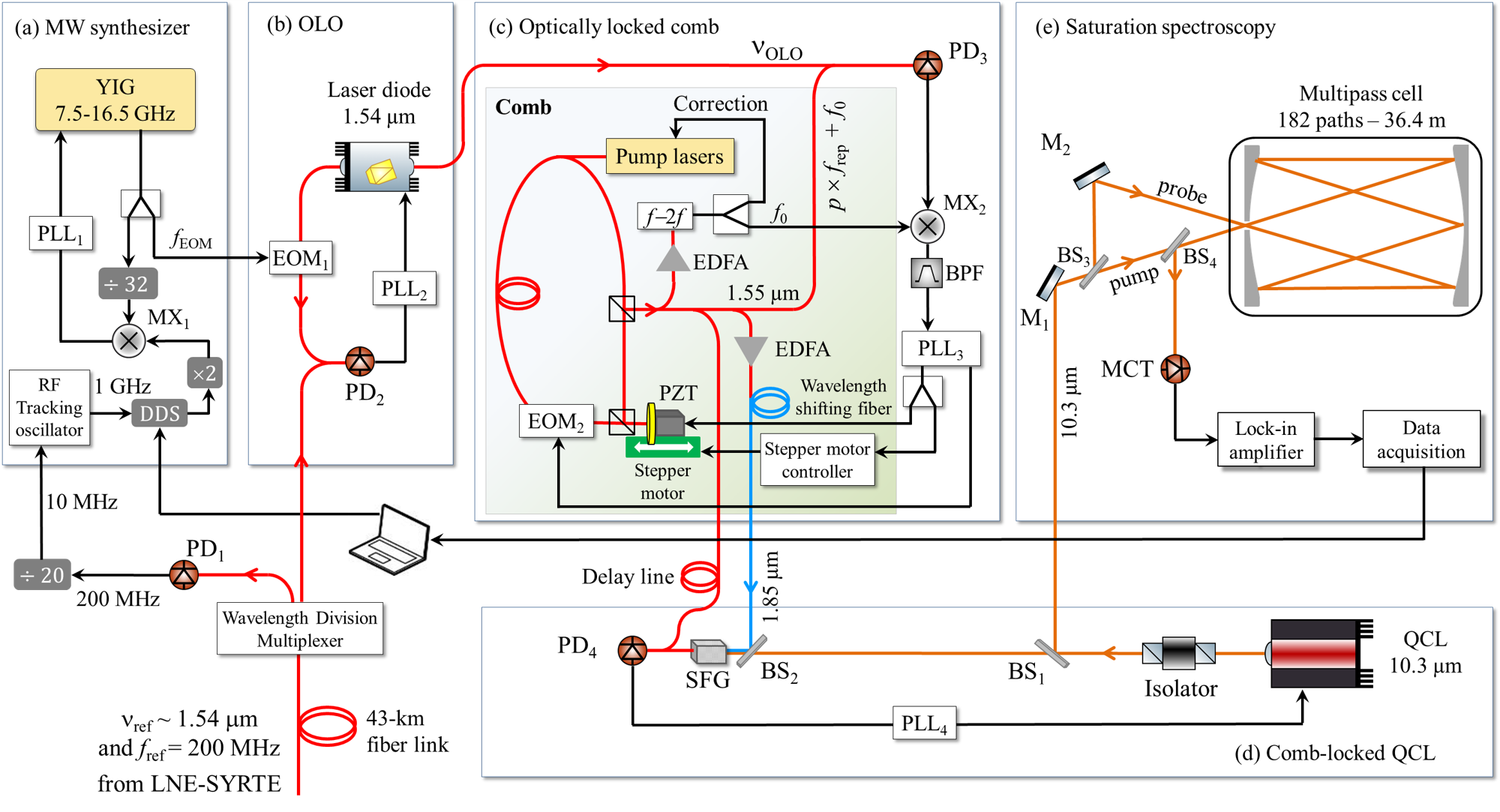}
\caption{Widely-tunable SI-traceable frequency-comb-stabilized-QCL-based high-resolution mid-IR spectrometer. (a) Yttrium Iron Garnet (YIG) oscillator-based microwave (MW) synthesizer disciplined to a 10 MHz RF reference. (b-d): QCL frequency stabilization to a remote 1.54 µm frequency reference using an optical frequency comb. The frequency reference operated at LNE-SYRTE is transferred to LPL via a 43-km long fiber link. In (b), a laser diode of frequency 1.54 µm is used as an optical local oscillator (OLO). One sideband generated by an electro-optic modulator (EOM$_1$) is locked to the remote reference using phase-lock loop PLL$_{2}$, ensuring wide tunability of the OLO carrier. In (c), the comb repetition rate $f_{\text{rep}}$ is locked to the OLO using PLL$_{3}$. In (d), a mid-IR quantum cascade laser (QCL) at 10.3 µm is stabilized to $f_{\text{rep}}$ using PLL$_{4}$. (e): Simplified setup for saturation spectroscopy in a multi-pass absorption cell using frequency modulation. DDS: direct digital synthesizer, MX$_\mathrm{i}$: mixers, PD$_\mathrm{i}$: photodiodes, $f-2f$: comb offset frequency detection, EDFA: Erbium-Dopped fibered amplifier, PBF: band pass filter, PZT: piezo-electric actuator, SFG: sum frequency generation crystal, BS$_\mathrm{i}$: beam splitters, M$_\mathrm{i}$: mirrors, MCT: mercury-cadmium-telluride photodetector.}
\label{fig_1}
\end{figure*}

The widely tunable OLO assembly is shown in FIG. \ref{fig_1}(b). The frequency reference $\nu_{\text{ref}}$ located at LNE-SYRTE is produced by a $\sim$1.54 µm fiber laser locked to an ultrastable cavity which exhibits a relative frequency stability lower than 10$^{-15}$ between 0.1 s and 10 s integration time \cite{santagata2019high,argence2012prototype}. This reference is calibrated against a combination of a liquid-helium cooled cryogenic sapphire oscillator (CSO) and a hydrogen maser (H-maser), themselves monitored on the atomic fountains of LNE-SYRTE, and its absolute frequency is thus SI-traceable to the primary standards with an uncertainty at the few 10$^{-14}$ level. This reference signal is transferred to LPL \cite{cantin2021accurate,xu2019reciprocity} through a 43-km long fiber link of the REFIMEVE infrastructure \cite{cantin2021accurate} without any degradation of its stability and absolute uncertainty thanks to an active compensation of the propagation-induced phase noise \cite{newbury2007coherent}. A 200 MHz ultrastable RF reference $f_{\text{ref}}$ derived from the CSO/H-maser combination ($\sim10^{-15}$ stability at 1 s, a few 10$^{-14}$ absolute frequency uncertainty \cite{guena2012progress}) is also transferred to LPL through the same 43-km long fiber link using an amplitude modulated auxiliary 1.5 µm laser diode.

At LPL, a laser diode of frequency $\nu_{OLO}$ $\sim$1.54 µm is used as an OLO. Two sidebands are generated in the OLO signal by an electro-optic modulator (EOM$_1$ in FIG. \ref{fig_1}(b), MPZ-LN-20, Photline). EOM$_1$ frequency $f_{\text{EOM$_1$}}$ is precisely controlled using a home-made phase-jump-free microwave synthesizer based on a current driven Yttrium Iron Garnet (YIG) oscillator with a continuous tuning range of 9 GHz (7.5-16.5 GHz) \cite{santagata2019high}. FIG. \ref{fig_1}(a) shows a simplified scheme of the synthesizer. A 1 GHz signal is generated by a RF tracking oscillator (TO) referenced to a 10 MHz signal (stability at 1 s and relative frequency uncertainty better than 10$^{-12}$ and 10$^{-13}$ respectively \cite{tran2019widely}) synthesized from the remote LNE-SYRTE 200 MHz reference $f_{\text{ref}}$. A 0-400 MHz direct digital synthesizer (DDS, AD9858, Analog Devices) uses the 1 GHz signal as a reference. After division by 32, the YIG frequency is mixed with the doubled DDS frequency. The resulting phase error signal is converted into correction signals (using phase-lock loop PLL$_{1}$ in FIG. \ref{fig_1}(a)) applied to the YIG oscillator’s current, the frequency of which is thus locked to the DDS with a frequency ratio of 64. This allows not only the spectral performance of the local 10 MHz reference to be transferred to the resulting microwave signal, but also the synthesizer frequency to be tuned over 9 GHz without any phase jumps by adjusting the DDS frequency. As illustrated in FIG. \ref{fig_1} and detailed below, the QCL is stabilized to the OLO via a series of three cascaded phase-lock loops. The wide tunability and, crucially, the phase-jump-free nature of the YIG-based synthesizer are key to allow broad but also continuous tunability of the stabilized QCL, and therefore demonstrate both broadband and ultra-high resolution spectroscopy.

The upper OLO sideband of frequency $\nu_{\text{OLO}}+f_{\text{EOM$_1$}}$ is phase-locked to the reference frequency $\nu_{\text{ref}}$ using PLL$_{2}$ (500 kHz locking bandwidth) in FIG. \ref{fig_1}(b) with an offset frequency $\Delta_{1}$. Adjusting $f_{\text{EOM$_1$}}$ then allows the OLO carrier frequency $\nu_{\text{OLO}}$  to be tuned over 9 GHz at the precision of the frequency reference $\nu_{\text{ref}}$ \cite{santagata2019high}. In addition to copying the 10$^{-15}$ stability (for 0.1 s to 10 s averaging times) of the LNE-SYRTE optical reference, the OLO carrier frequency can be used as a tunable SI-traceable local oscillator of frequency:
\begin{equation}
    \label{eq_1}
    \nu_{\text{OLO}}=\nu_{\text{ref}}-f_{\text{EOM$_1$}}-\Delta_{1}.
\end{equation}

\subsection{\label{sec2_2}Widely tunable, SI-traceable and ultrastable frequency comb-assisted quantum cascade laser}

A $f_{\text{rep}}$ $\sim$250 MHz repetition rate Erbium-doped fiber optical frequency comb (Menlo Systems, F1500) is used to transfer the spectral performances of the remote frequency reference and the wide-tunability of the OLO to the QCL.

A beat-note signal at 
\begin{equation}
    \label{eq_2}
    \Delta_{2}=\nu_{\text{OLO}}-p \times f_{\text{rep}}
\end{equation}
is obtained after beating the OLO carrier frequency with the $p^{th}$ nearest comb tooth at frequency $p\times f_{\text{rep}} +f_{\text{0}}$  (with $p \sim780\;000$) and subsequently removing the comb offset frequency ($f_{\text{0}}$). This signal is used to lock $f_{\text{rep}}$. To this end, it is processed via PLL$_{3}$ to generate three correction signals respectively applied to a stepper motor (very slow correction), a piezo-electric actuator (PZT, slow-correction) and an intra-cavity electro-optic modulator EOM$_2$ (fast-correction), all used to act on the comb  cavity optical length (see panel (c) in FIG. \ref{fig_1}, the stepper motor and the PZT are both mounted on a cavity mirror). Phase-locking $f_{\text{rep}}$ also involves transferring the tunability of the OLO carrier to every comb modes. Here, a couple of improvements have been made to the  previous setup \cite{santagata2019high}. The dynamic range of EOM$_2$ control voltage has been doubled, resulting in a larger locking bandwidth of $\sim$700 kHz. The most significant upgrade lies in the use of the stepper motor in order to enhance the tuning range of the cavity length and thus of $f_{\text{rep}}$ (limited in the previous setup by the course of the PZT actuator, corresponding to a 3 GHz span of a comb mode). To prevent the PZT from reaching the end of its course in our new setup, its voltage is maintained within a reduced range around its mid-point by acting on the stepper motor. When the PZT voltage falls outside the allowed range, a correction signal is sent to the stepper motor which performs a translation that brings the PZT back to its mid-point. This allows to lock $f_{\text{rep}}$ and tune it over a span corresponding to a 9 GHz tunability of the comb modes. This span is 3 times broader than allowed by acting on the PZT only, and now limited by EOM$_1$ tuning range, given by the YIG span. Sweeping $f_{\text{rep}}$ over such extended spans however also involves variations of $f_{0}$ of several tens of megahertz, thus pushing some beat notes (not only $f_{\text{0}}$ itself but also the beat signal between the OLO and the $p^{th}$ comb tooth) outside filters’ bandwidths. The comb offset frequency is thus loosely locked (sub-10 Hz bandwidth) by acting on the comb pump lasers’ currents. 

We then lock the QCL (frequency $\nu_{\text{QCL}}$) to a high harmonic of $f_{\text{rep}}$  (FIG. \ref{fig_1}(d)). To this end, a fraction of the 1.54 µm frequency comb power is amplified in an Erbium-doped fiber amplifier (EDFA) and fed to a wavelength  shifting fiber (WSF) to generate a new comb output centered at 1.85 µm with comb teeth frequencies $q\times f_{\text{rep}} +f_{0}$ ($q \sim650\;000$). The QCL and 1.85 µm comb beams are overlapped in a non-linear crystal (AgGaSe$_2$) to perform sum frequency generation (SFG) which yields a shifted comb centered at 1.55 µm (SFG comb). The beat signal between the SFG and original combs at frequency:
\begin{equation}
    \label{eq_3}
    \Delta_{3}=n\times f_{\text{rep}} - \nu_{\text{QCL}}
\end{equation}
with $n=p-q \sim120\;000$ is used to lock the QCL frequency. It is obtained after overlapping the pulses in the time domain using a fiber delay line (DL in FIG. \ref{fig_1}). It is processed by PLL$_{4}$ to generate a correction signal applied to the QCL’s current. The QCL frequency is thus directly traceable to the remote frequency reference in the following way:
\begin{equation}
    \label{eq_4}
    \nu_{\text{QCL}}=\frac{n}{p}\left( \nu_{\text{ref}}-f_{\text{EOM$_1$}} - \Delta_{1}-\Delta_{2}\right)-\Delta_{3}
\end{equation}
with $n/p \sim0.15$. Integers $n$ and $p$ are unequivocally determined by measuring the comb repetition rate using a RF counter (K\&K FXE, dead-time-free), the QCL and OLO frequencies using an optical spectrum analyzer (Bristol Instruments, model 771A-MIR, 0.2 ppm accuracy), and by exploiting Equations (\ref{eq_2}) and (\ref{eq_3}) \cite{santagata2019high}.

As demonstrated in Ref. \cite{santagata2019high,argence2015quantum}, the QCL frequency (i) exhibits a stability at the level of the remote reference signal, below 0.03 Hz ($10^{-15}$ in relative value) for averaging times from 0.1 to 10 s, and a linewidth of $\delta\nu\sim$0.1~Hz at 100 ms timescales; (ii) is SI-traceable and known within a total uncertainty better than $4\times 10^{-14}$ after 1 s averaging time. 

Scanning the OLO frequency thus allows us to continuously tune $f_{\text{rep}}$ and in turn the QCL frequency at the precision of the remote reference. In a previous work \cite{santagata2019high}, we have demonstrated a QCL continuous tunability over a span of $\sim$400 MHz, limited by the course of the PZT actuator. With the additional capability provided by the stepper motor control, scanning the OLO carrier frequency over its entire 9 GHz span (limited by the YIG tunability) enables the QCL, which is both ultrastable and SI-traceable, to be continuously tuned over a range of 1.4 GHz (given by the $\sim$6.7 mid-IR to near-IR wavelength ratio). However, as explained in Appendix \ref{app_E}, if the time delay between the SFG and original combs is not cancelled, tuning $f_{\text{rep}}$ over the extended spans resulting from the use of the stepper motor comes with a deterioration of the pulses time overlap and in turn of beat-note signal $\Delta_{3}$'s signal-to-noise ratio, preventing us from carrying out long scans. Time delay cancellation is achieved by adding fiber length in order to balance the optical paths (see details in Appendix \ref{app_E}). The combination of tunability and spectral purity in our spectrometer yields a $\Delta\nu/\delta\nu\sim10^{10}$ continuous span, setting to our knowledge a new record both for the mid-IR QCL and the near-IR comb teeth.

\subsection{\label{sec2_3}Saturated absorption spectroscopy setup}
As illustrated in FIG. \ref{fig_1}(e), the stabilized QCL has been used to carry out saturated absorption spectroscopic measurements in a multi-pass absorption cell (Aerodyne Research, model AMAC-36, 182 passes). With a 20-cm long distance between two astigmatic mirrors, the cell provides an effective absorption length of 36.4 m and allows us to perform spectroscopic measurements of weak lines at low pressures.

The QCL beam is split in two using an 80/20 beam splitter (BS$_{1}$). Around 8 mW is needed to phase-lock the laser to the comb while about 2 mW remains for the spectroscopy. Using BS$_{2}$, the beam is further split into a pump and a probe beam, which are coupled into the multi-pass cell (incident powers of $\sim$1.3 mW and $\sim$0.7 mW, respectively) for conducting saturation spectroscopy. The mirrors’ reflectivity leads to a $\sim$20\% transmission after 182 passes which in turn results in pump and probe powers that vary inversely by almost an order of magnitude through the cell. This also goes with a $\sim$50\% total power (sum of pump and probe powers) variation from pass to pass and to an intra-cell total averaged power of 0.81 mW (averaged over the 182 passes). After exiting the cell, the probe beam is detected by a liquid-nitrogen-cooled mercury-cadmium-telluride photodetector (MCT in FIG. \ref{fig_1}(e)). Undesirable interference fringes as typically observed with multi-pass cells \cite{mcmanus1995astigmatic} are averaged out by vibrating mirror M$_{2}$ held in a piezo-actuated mount and shaking the multi-pass cell assembly with a fan. 

Frequency modulation (FM) of the QCL is used to improve the signal-to-noise ratio. This is done by modulating the frequency of the DDS used as a reference for PLL$_{4}$ (to lock the QCL to the comb) at a frequency of 20 kHz (well within PLL$_{4}$’s 250 kHz bandwidth). The signal detected by the MCT photodetector is fed to a lock-in amplifier for demodulation.

\section{PRECISE SPECTROSCOPY OF METHANOL}
Saturated absorption spectroscopy of methanol (purchased from VWR Chemicals, $\geq99.8$\% purity, see Appendix~\ref{app_AA} for more details on the sample purity) is carried out by tuning the QCL frequency in a series of discrete steps. To cancel out the frequency shifts induced by the limited detection bandwidth \cite{rohart2014absorption,rohart2017overcoming} we have carried out frequency scans in both directions with increasing and decreasing frequencies and only consider the averaged spectrum of pairs of up and down scans (see below). FIG. \ref{fig_2}(a) shows a saturated absorption spectrum of  methanol averaged over five such pairs spanning the full 1.4 GHz tuning range of our system (from $\sim$971.312 to $\sim$971.357 cm$^{-1}$) after demodulation on the first harmonic. It has been recorded at a pressure of 1.5 Pa with a frequency step $\sim$15 kHz and a step duration and lock-in amplifier time constant both of 10 ms. The SI-traceability detailed in Section \ref{sec2_2} allows us to retrieve the absolute frequency scale using Equation (\ref{eq_4}). Although $\nu_{\text{ref}}$ is measured and made available every second at LNE-SYRTE, its value varies by  a few hertz at most over the duration of our scans. Therefore, for each pairs of scans, we fix $\nu_{\text{ref}}$ in Equation (\ref{eq_4}) to its value measured halfway through the scans.

\begin{figure*}
\includegraphics[width=0.9\textwidth]{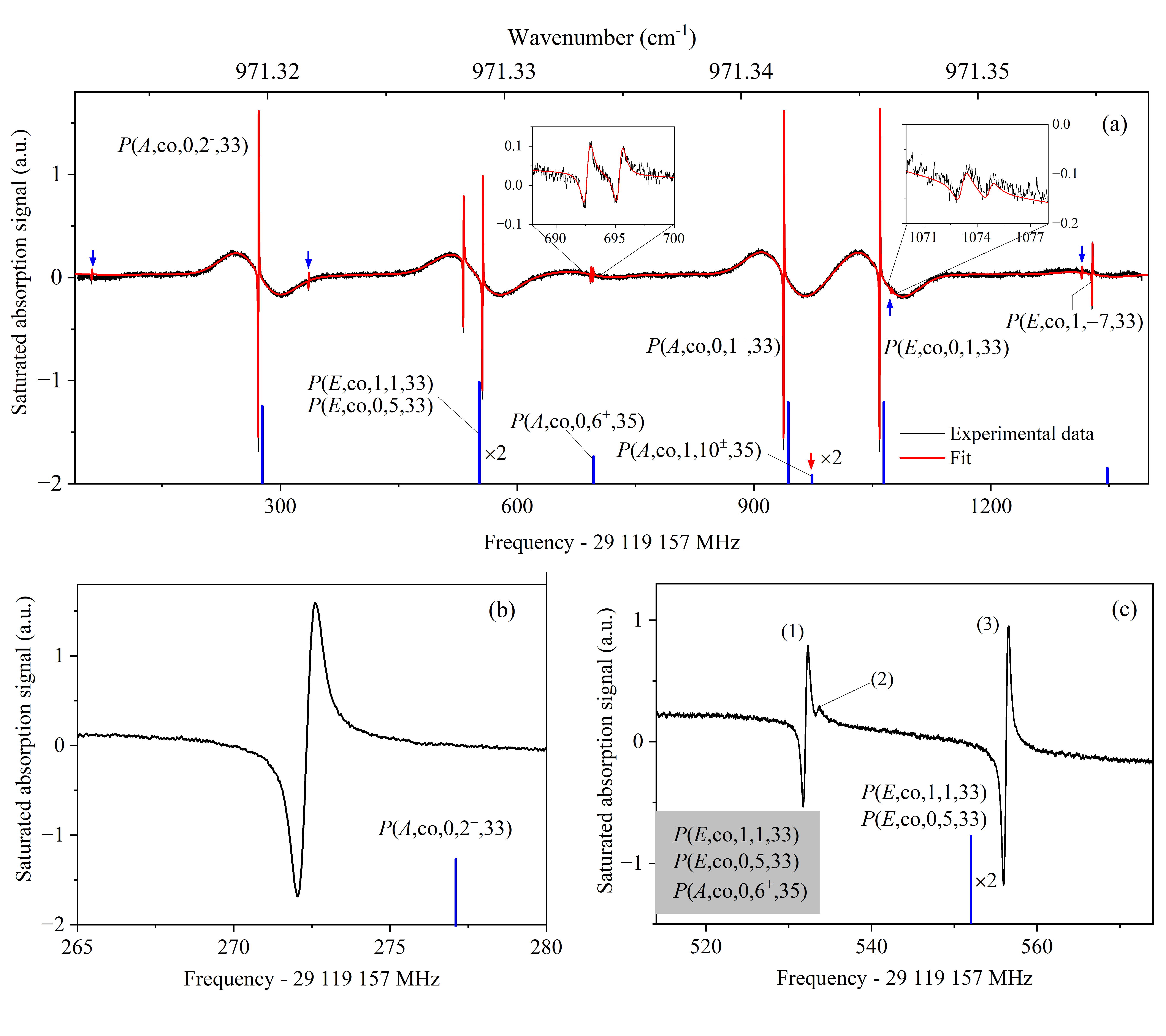}
\caption{(a) Saturated absorption spectrum of methanol spanning $\sim$1.4 GHz recorded using frequency modulation and first-harmonic detection (black curve). Blue sticks indicate the line frequencies and intensities reported in the HITRAN database (intensities of degenerate lines are summed). Experimental conditions: pressure, 1.5 Pa; modulation frequency, 20 kHz; modulation amplitude, 250 kHz; frequency step, $\sim$15 kHz; average of five pairs of up and down scans; total integration time per point, 100 ms; whole spectrum measurement time, 9000 s. Transitions marked with blue arrows correspond to weak transitions that we have not been able to assign (see text). (b) Zoom on the saturated absorption spectrum of the $P(A,\mathrm{co},0,2^-,33)$ rovibrational transition. (c) Zoom of the saturated absorption features observed around 971.328 cm$^{-1}$. Lines (1-3) are tentatively assigned to the transitions listed in the grey rectangle.}
\label{fig_2}
\end{figure*}

The spectrum exhibits fourteen rovibrational transitions of methanol which belong to the $P$ branch of the $\nu_{8}$ C-O stretch vibrational mode \cite{xu2004new}. The red solid line is a fit to the data used as a guide-to-the-eye. Each line is fitted with the first derivative of the sum of a Lorentzian and a Gaussian profile to model the saturated absorption and Doppler (oscillations in the baseline) contribution, respectively. The transitions reported in the HITRAN database \cite{gordon2022hitran2020} falling in the spectral window covered are shown as blue sticks. As an example, FIG. \ref{fig_2}(b) shows a zoom on the saturation feature of the $P(A,\mathrm{co},t=0,K=2^{-},J=33)$ methanol line around 971.319~cm$^{-1}$ (in this work, we adopt the notations of Ref. \cite{xu2004new} for the spectroscopic assignment of methanol transitions, see also Appendix \ref{app_A}). It exhibits a signal-to-noise ratio of $\sim$370 and a $\sim$760 kHz full-width-at-half maximum of the free-of-FM-induced-distortion Lorentzian underlying the line shape. The latter is a combination of transit time ($\sim120$~kHz) pressure ($\sim330$~kHz) as well as power broadening.

To achieve a reasonable signal-to-noise ratio, the FM amplitude has been set at 250 kHz, resulting in a line shape slightly distorted compared to the typical Lorentzian profile. Furthermore, the residual amplitude modulation associated with FM and the power variation over a scan can both contribute to an asymmetry of the line shape. We fit our data to a model described in Appendix \ref{app_B} that takes into account these complex distortions. It is based on a direct absorption profile introduced by Arndt~\cite{arndt1965analytical} and Schilt \emph{et al}~\cite{schilt2003wavelength} that we have extended to the analysis of saturation spectra. Exploiting this model has allowed us to use larger FM amplitudes directly resulting in an increase in signal-to-noise ratio.

\begin{table*}
	\caption{  Line-center frequencies of rovibrational transitions of methanol shown in FIG. \ref{fig_2}, (tentative) assignments and corresponding values reported in the HITRAN database\cite{gordon2022hitran2020} and in Ref. \cite{lees2007rotation} when a match is possible.}
	\label{table_1}
	\centering
	\setlength{\extrarowheight}{3pt}%
	\begin{ruledtabular}
		\begin{tabular}{p{3.5cm}p{2.5cm}p{1.8cm}p{1.8cm}p{2.0cm}p{2.2cm}p{1.8cm}p{1.8cm}}
			\centering This work & \centering\multirow{2}{*}{Assignment}	&\multicolumn{2}{c}{HITRAN}	&  \centering Ref.  \cite{lees2007rotation} & $\nu_{\text{TW}}-\nu_{\text{HITRAN}}$ &$\nu_{\text{TW}}-\nu_{\text{Ref. \cite{lees2007rotation}}}$ \\
		\cline{3-4}
			
		\centering($\nu_{\text{TW}}$, kHz)	&& \centering Wavenumber (cm$^{-1}$)	& \centering Intensity (cm$^{-1}$/mol.cm$^{-2}$)& \centering Wavenumber (cm$^{-1}$) & \hspace{0.5cm} (MHz)  & \hspace{0.3cm} (MHz)\\
			\hline
		29 119 218 330 (120)  &	$-$                   &$-$        &$-$                  &$-$   & &\\	
			\hline
		29 119 429 315.1 (8.6)    &	$P(A,\text{co},0,2^{-},33)$  &971.31977  &$5.950\times 10^{-22}$ & \hspace{0.3cm} 971.31966 &$-4.8$ &$-1.5$ \\
			\hline
		29 119 492 940 (130)  &	$-$			          &$-$        &$-$              & $-$      & &\\
			\hline
		29 119 689 047.3 (7.1)     &  $P(E,\text{co},1,1,33)^{\text{(a)}}$ 
      &
  \multirow{2}{*}{\hspace{-0.7cm}$\left. \begin{array}{l}
				\\
				\\
				\end{array}
				\right\}$ \hspace{0.05cm} 971.32894}                                  &$4.712\times 10^{-22}$ & \multirow{3}{*}{\hspace{-0.6cm}$\left. \begin{array}{l}
				\\
				\\
				\\
				\end{array}
				\right\}$ \hspace{0.3cm} 971.32880} & &\\
		29 119 690 401 (39) &$P(E,\text{co},0,5,33)^{\text{(a)}}$   &           &$3.083\times 10^{-22}$& & &\\
		29 119 713 237 (11)   &$P(A,\text{co},0,6^{+},35)^{\text{(a)}}$  &971.33377  & $2.076\times 10^{-22}$ & &  & \\	

			\hline
		29 119 850 247 (81)$^{\text{(b)}}$  &\multirow{2}{*}{$P(A,\text{co},1,10^{\pm},35)^{\text{(c)}}$ }& 971.34300  &$3.215\times 10^{-23}$ & & &\\
		29 119 853 047 (74)$^{\text{(b)}}$	  &   &971.34300	&$3.215\times 10^{-23}$ & &	&\\
			\hline
		29 120 094 739.6 (6.2)	  &$P(A,\text{co},0,1^{-},33)$	  &971.34199  &$6.251\times 10^{-22}$ & \hspace{0.3cm} 971.34176 &$-5.5$ & $+1.4$\\
			\hline
		29 120 216 332.4 (7.5)    &$P(E,\text{co},0,1,33)$		 &971.34604   &$6.263\times 10^{-22}$ & \hspace{0.3cm} 971.34592 &$-5.4$&$-1.8$\\
			\hline
		29 120 230 693 (75)$^{\text{(b), (d)}}$  & $-$ &$-$&$-$ &$-$  & &\\
		29 120 232 360 (170)$^{\text{(b), (d)}}$  &$-$	&$-$&$-$&$-$  & &\\		
			\hline
		29 120 472 397 (86)   &$-$			         &$-$         &$-$              &  $-$      & 	&\\
			\hline
		29 120 485 592 (26)  &$P(E,\text{co},1,-7,33)$	 &971.35548   &$1.181\times 10^{-22}$ & \hspace{0.3cm} 971.35536 &$-19.1$ &$-15.5$\\	
			
		\end{tabular}
	\end{ruledtabular}
	\begin{flushleft}
		\footnotesize{
			$^{\text{(a)}}$Tentative assignments of the three measured and resolved lines without being able to decide which one is which.\\
                $^{\text{(b)}}$Measured resolved doublets.\\
                $^{\text{(c)}}$Tentative assignments.\\
                $^{\text{(d)}}$Only four average pairs of up and down scans have been used for these lines.\\
		}	
	\end{flushleft}
\end{table*} 

To determine line-center frequencies of the fourteen resolved transitions, we first select a spectral range of $\sim$6 MHz around each. We perform the following “pair by pair” analysis already established in \cite{santagata2019high}. We average each pair of up and down scans resulting in five averaged spectra for each of the fourteen transitions. To all data points of an averaged spectrum, we assign the same experimental error bar, calculated as the standard deviation of the residuals obtained after fitting a second-order polynomial to a small portion of the averaged spectrum far from resonance. To each of the five averaged spectra of a given transition is fitted the sum of the line profile described in Appendix \ref{app_B} and of a second order polynomial to account for the baseline (see Appendix \ref{app_C}). The absolute frequency and associated statistical uncertainty of each line are estimated by calculating the weighted mean and weighted standard error of the five fitted center frequencies, with the weights determined from the fits’ error bars. Unlike in our previous work \cite{santagata2019high}, we have conducted measurements at a single pressure (1.5 Pa) and power (0.81 mW intra-cell averaged power corresponding to a saturation parameter slightly larger than 1 for the most intense lines probed) preventing us from deducing zero-power and collision-free transition frequencies. However, we estimate the resulting overall pressure- and power-shift to be of the order of 10 kHz. The reader is referred to Refs. \cite{santagata2019high,tran2019widely} and to the summary given in Appendix \ref{app_D} for a detailed description of the line positions uncertainty budget which result in a systematic uncertainty of 5.4 kHz on the frequency of all the resonances studied here. TABLE \ref{table_1} lists the line center frequencies of the fourteen transitions recorded at a pressure of 1.5 Pa and an intra-cell averaged power of 0.81 mW. 1-$\sigma$ total uncertainties are quoted into parentheses. The absolute frequencies of three unambiguously assigned high intensity transitions have been determined with a sub-10-kHz total uncertainty ($2.1\times10^{-10}$ at best in relative value), an improvement of $\sim$2000 over previous measurements based on the Fourier-transform IR (FTIR) spectroscopic technique \cite{xu2004new,gordon2022hitran2020}. All other absolute frequencies are determined with a total uncertainty ranging from $\sim$10 kHz to $\sim$350 kHz, to be compared to the typical 15 MHz FTIR uncertainty \cite{xu2004new}.

The data shown in FIG. \ref{fig_2} demonstrates both the high detection-sensitivity and high resolution of our spectrometer. An illustration of this double asset is the resolution of two weak doublets (see insets of FIG. \ref{fig_2}(a)), which exhibit a splitting of $2.66 \pm 0.07$ and $1.70 \pm 0.16$ MHz, respectively. The use of a multi-pass cell allows relatively high-$J$ $(J>30)$ rovibrational lines to be probed at the low pressures required for ultra-high resolution measurements, including weak lines belonging to hot-bands (torsional excitation $t \geqslant 1$). We have thus been able to demonstrate sub-10 kHz uncertainties for the strongest transitions shown here exhibiting intensities of several $10^{-22}$~ cm$^{-1}$/(mol.cm$^{-2}$) two to three orders of magnitude weaker than the most intense lines of this vibrational mode, and $\sim100$~kHz uncertainties for lines as weak as a few $10^{-23}$~cm$^{-1}$/mol.cm$^{-2}$.

As shown in FIG. \ref{fig_2}(a), our spectrum is a very rich source of information. First of all, we resolve fourteen lines where the HITRAN database lists only nine (of which two pairs reported as degenerate). Several transitions have to our knowledge never been reported elsewhere. Based on HITRAN, we are in fact able to unequivocally assign only four of the lines (see TABLE \ref{table_1}). Five weak transitions (including one doublet) that could not be assigned are indicated with blue arrows and most probably belong to hot-bands with torsional excitation $t \geqslant 2$. As discussed in the following and summarized in TABLE \ref{table_1}, the five remaining measured resonances are tentatively assigned to the five other HITRAN transitions. As shown in FIG. \ref{fig_2}(c), around 971.328 cm$^{-1}$ we resolve one weak and two intense lines labelled (1), (2) and (3). We tentatively assign these to transitions $P(E,\mathrm{co},1,1,33)$, $P(E,\mathrm{co},0,5,33)$ and $P(A,\mathrm{co},0,6^+,35)$, without being able to decide which one is which. Those three transitions are indeed listed in Ref. \cite{lees2007rotation} at the same degenerate frequency of 971.3288 cm$^{-1}$. In HITRAN however and as shown in FIG.~\ref{fig_2}, only $P(E,\mathrm{co},1,1,33)$ and $P(E,\mathrm{co},0,5,33)$ are reported as degenerate at 971.32894~cm$^{-1}$, while $P(A,\mathrm{co},0,6^+,35)$ is $\sim$150 MHz blue-shifted at 971.33377 cm$^{-1}$. It is however unlikely for this latter to correspond to one component of the doublet observed at $\sim$971.334 cm$^{-1}$ (based on intensities, this doublet looks very much like a single structure, not like two fortuitously quasi-coincident lines). $P(A,\mathrm{co},1,K=10^{\pm},35)$ (red arrow in FIG. \ref{fig_2}(a)) corresponds to a so-called closed-lying K-doublet that splits for $A$-symmetry lines due to the combination of the slight asymmetry and internal tunnel dynamics of the methanol molecule~\cite{lees2007rotation}.  It is reported as degenerate in the HITRAN database, but no matching resonance is observed around. Even at the high resolution demonstrated here, we do not expect to resolve such $K$-doublets for $K\geqslant7$ (here $K = 10$), but enhanced splittings are possible as a result of mixing of the considered transition upper states that belong to the C-O stretching vibrational mode with a closely lying state from another vibrational mode \cite{lees2007rotation}. $P(A,\mathrm{co},1,K=10^{\pm},35)$ could then well be assigned to either of the two observed resolved doublets. Based on intensities, we tentatively assign it to the strongest doublet around 971.334 cm$^{-1}$. All certain and tentative assignments are finally summarized in TABLE \ref{table_1}.

Our data is a source of information much richer than what is currently available in the literature. It has yet to be fully exploited to build a more accurate spectroscopic model of methanol. As seen, it brings to light some inaccuracies and gaps in the HITRAN database. In addition, our measured and unequivocally assigned resonance frequencies are red-shifted by 5 to 20 MHz with respect to previous FTIR measurements at the origin of the current HITRAN edition line list \cite{gordon2022hitran2020} (see FIG. \ref{fig_2} and TABLE \ref{table_1}). These shifts are about three orders of magnitude larger than our uncertainties and more than an order of magnitude larger than the HITRAN frequency accuracies reported to be between 30 kHz and 300~kHz. It is consistent with deviations between FTIR and saturation spectroscopy data previously observed in the C-O stretch of methanol  and is attributed to FTIR spectrometers calibration imperfections~\cite{santagata2019high,sun2000sub,sun2002sub,sun2003sub}. Our data also offers new information on weak hot-band transitions for which molecular databases remain largely incomplete. Exploiting this in refined models of methanol is essential for atmospheric quantification. Moreover, mixing between near-degenerate levels of two different modes is known to lead to collision-induced population transfer from one vibrational mode to another. Doublets exhibiting enhanced splittings such as those resolved in our work, a signature of this type of mixing, may thus help understand how molecules transfer among different modes and give insight on  thermal equilibration in gases \cite{lees2007rotation}.

Other saturation spectroscopy measurements of the C-O stretch of methanol exist. Almost 700 frequencies have been measured with an accuracy of $\sim$100 kHz using a CO$_{2}$-laser spectrometer \cite{sun2000sub,sun2002sub,sun2003sub,sun2004dual,sun2006saturation}, however mostly in the $Q$ and $R$ branches, with a few low $J$ lines in the $P$ branch. To our knowledge, there are only two other methanol frequencies that have been measured with an uncertainty comparable to the present work: our previous measurement \cite{santagata2019high}, and a weak unassigned line~\cite{tochitsky1998precise} around 947.7 cm$^{-1}$.

\section{CONCLUSION}

We present our progress in the development and operation of a widely tunable SI-traceable frequency-comb-stabilized high-resolution spectrometer potentially covering the 8-12 µm spectral window. A mode-locked frequency comb and a metrological fiber link are used to transfer the spectral purity and SI-traceability of a 1.54 µm frequency reference located at LNE-SYRTE to a 10.3 µm QCL located at LPL. The QCL exhibits a sub-$10^{-15}$ relative frequency stability from 0.1 to 10 s, a linewidth of $\sim$0.1 Hz at 100 ms and its frequency is SI-traceable with a total relative uncertainty better than $4\times10^{-14}$ after 1 s of averaging time. In addition, we have developed the instrumentation allowing comb modes to be continuously tuned over 9 GHz, resulting in a continuous tunability of 1.4 GHz for the QCL, a more than three-fold improvement compared to previous measurements at such levels of spectral purity. This tuning range can potentially be increased by transferring the full 9 GHz tunability of comb modes directly to the QCL, or by using a commercially available 40 GHz telecom electro-optic modulator together with a home-made microwave synthesizer of wider tunability , giving potentially access to a continuous span of $\Delta\nu/\delta\nu > 10^{11}$. 

We have carried out saturation spectroscopy of methanol in a multi-pass absorption cell. We report line-center frequencies of fourteen transitions of methanol in the $P$ branch of the $\nu_{8}$ C-O stretch vibrational mode, including very weak transitions belonging to hot-bands, and some observed for the first time. We demonstrate record global uncertainties ranging from few kilohertz to $\sim100$~kHz – three to four orders of magnitude improved over previous measurements – depending on the line intensity. We expose manifest discrepancies with the HITRAN database, provide information on hot-bands essential to atmospheric quantification and resolve subtle weak intensity doublets never observed before induced by a combination of the asymmetry and the tunneling dynamics in methanol. The $10^{-10}$ level frequency accuracy achieved here which surpasses traditional high resolution Fourier transform spectrometers by 3 orders of magnitude thus allows molecular databases to be filled with increasingly accurate parameters. This is essential for improving spectroscopic models of molecules of atmospheric or astrophysical interest such as methanol. In this context, sub-Doppler spectroscopy potentially gives access to line positions, transition dipole moments, pressure broadenings and shifts as well as corresponding beyond-the-Voigt speed dependent parameters. Conducting saturation spectroscopy in a Fabry-Perot cavity rather than in a multi-pass cell using our sub-Hz frequency-comb-stabilized QCL would result in improved mid-IR measurements in methanol with improved uncertainties to the $10^{-11}$-$10^{-12}$ level~\cite{argence2015quantum}. This is important for (i) refining constraints on the variation of the proton-to-electron mass ratio µ by comparing laboratory data to spectra of cosmic objects~\cite{muller2021study}; (ii) unraveling the hyperfine structure in view of using methanol as a magnetic field tracer in star-forming regions~\cite{lankhaar_characterization_2018}. Besides, the sub-$10^{-15}$ level of mid-IR laser spectral purity achieved is important for pushing back the limits in ultra-high resolution molecular spectroscopy. It could be exploited to reach $10^{-12}$-$10^{-13}$ mid-IR frequency uncertainties on simple and calculable trapped hydrogen molecular isotopologues (H$_2^+$, HD$^+$, D$_2^+$,...) for providing stringent tests of quantum electrodynamics or precise measurements of µ, or to search for physics beyond the standard model by looking for fifth forces or extra dimensions at the molecular scale~\cite{patra2020proton,kortunov_protonelectron_2021}. Finally, the mid-IR spectral performance reported in the present work is mandatory for reaching uncertainty levels at the $10^{-14}$-$10^{-15}$ level or beyond, \emph{i.e.} required to test the parity symmetry by measuring the tiny energy differences between chiral enantiomers~\cite{Tokunaga2017,fiechter2022toward,segal2018studying} or to refine direct and model-free constraints on possible current-epoch variations of µ~\cite{shelkovnikov2008stability,barontini2022measuring,segal2018studying}, for which methanol is a promising candidate molecule.

\begin{acknowledgments}
This project has received funding from the EMPIR programme co-financed by the Participating States and from the European Union's Horizon 2020 research and innovation programme, through EMPIR project 15SIB05 “OFTEN”. This work was supported by the Region Île-de-France in the framework of DIM Nano-K and DIM SIRTEQ, the Agence Nationale de la Recherche project PVCM (Grant No. ANR-15-CE30-0005-01), the LabEx Cluster of Excellence FIRST-TF (ANR-10-LABX-48-01), the EquipEx Cluster of Excellence REFIMEVE+ (ANR-11-EQPX-0039), CNRS and Université Sorbonne Paris Nord. D.B.A. Tran was supported by the Ministry of Education and Training, Vietnam (Program 911).

\end{acknowledgments}

\section*{Data Availability}
The data that support the findings of this study are available from the corresponding author upon reasonable request.

\appendix

\section{Measurement and compensation of the time delay between the SFG and the  original comb}\label{app_E}

The wide tuning range of the stabilised QCL may be altered if the time delay between the SFG and original comb is non-zero, as a result of a propagation length difference $\Delta L$. Even if, for a given value $f_{\text{rep$_0$}}$ of the repetition rate, the pulses of the two combs are well time-overlapped  by using a short delay line (few tens of nanoseconds), there can exist a delay between combs $\Delta T=n\Delta L/c=m/f_{\text{rep$_0$}}$ corresponding to an integer number $m$ of the time between two adjacent pulses, \emph{i.e.} when the $k^{\text{th}}$ pulse of the SFG comb, with $k$ an integer, overlaps with the $(k+m)^{\text{th}}$ pulse of the original comb (we assume here propagation in a medium of refractive index $n$, with $c$ the speed of light). Tuning $f_{\text{rep}}$ away from $f_{\text{rep$_0$}}$ results in a gradual loss of the overlap and in a time-separation between SFG and original comb pulses of $\delta_{\text{pulse}}\simeq m\delta f_{\text{rep}}/f_{\text{rep$_0$}}^2$ for small variations $\delta f_{\text{rep}}$ of the repetition rate. To avoid degrading the signal-to-noise ratio of the two-comb beat-note used to lock the QCL, $\delta_{\text{pulse}}$ should remain smaller than the combs pulse width (a few hundreds of femtoseconds) during a scan.

\begin{figure}[h]
	\centering
	\includegraphics[scale=0.4]{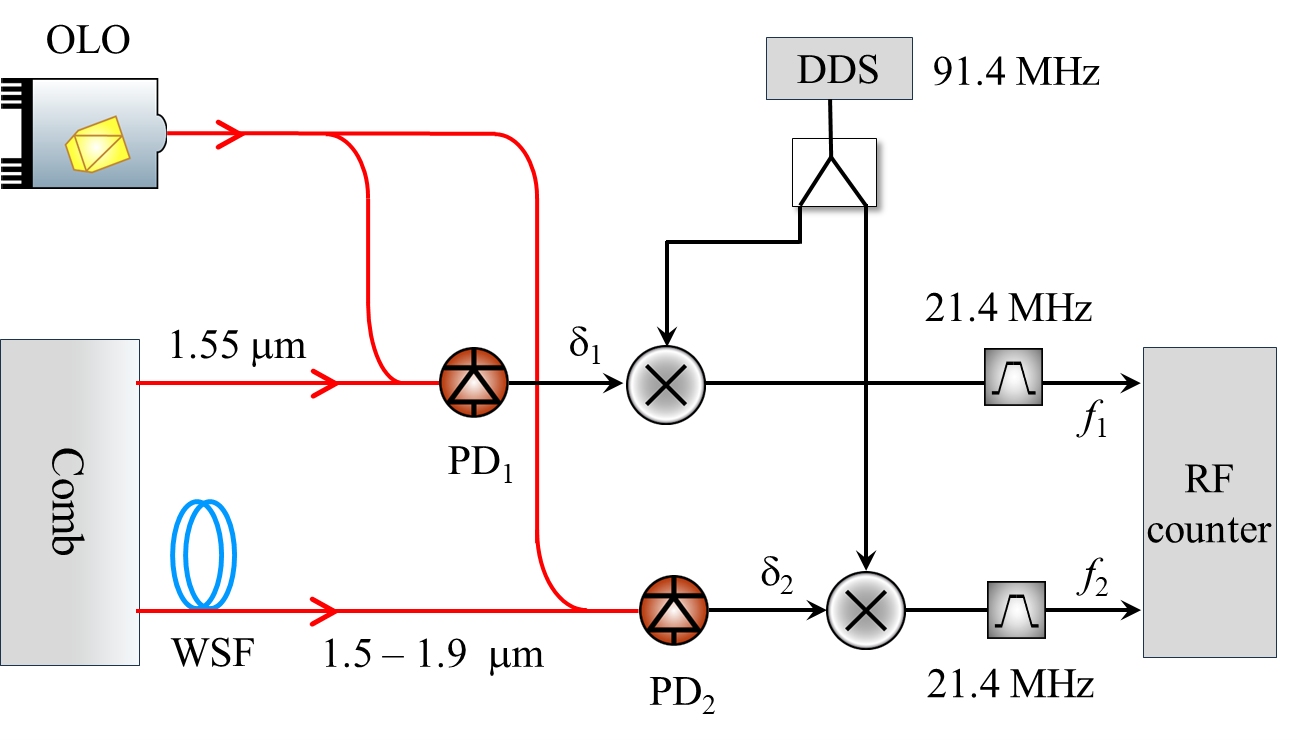}
	\caption{ Experimental setup for measuring the time-separation $\delta_{\text{pulse}}$ between the pulses of the original comb and the additional output (after the wavelength shifting fiber) used to generate the SFG comb. The comb repetition rate is phase-locked onto the OLO frequency. Two beat-note signals around 70 MHz are detected: $\delta_{1}$ between a tooth of the original comb and the OLO on photodiode PD$_1$ and $\delta_{2}$ between the same tooth of the additional comb and the OLO on photodiode PD$_2$. Their frequencies are converted to $f_1(t)$ and $f_2(t)$ at $\sim$21.4 MHz by mixing with a radio-frequency signal at 91.4 MHz and counted by a counter. OLO: optical local oscillator, RF: radio-frequency, DDS: direct digital synthesizer.}
	\label{figA2_1}
\end{figure}

As shown in Fig.~\ref{figA2_1}, we have set up an experiment to measure the delay between the pulses of the original comb and the additional output (after the wavelength shifting fiber) used to generate the SFG comb. We lock $f_{\text{rep}}$ to the OLO as detailed in the main text and we detect the two beat-note signals at frequency $f_1(t)$ and $f_2(t)$ between the OLO and the closest $p^{\text{th}}$ tooth of the original, respectively additional comb. A $\Pi$-type RF frequency counter is used to count $f_1(t)$and $f_2(t)$ (1~ms gate time) and the frequency difference $\Delta f(t)=f_{2}(t)-f_{1}(t)$ is processed. If ${\nu_{\text{OLO}}}$ is kept fixed, $\Delta f(t)=0$ ($f_{\text{rep}}$ is the same for both combs). During a scan, however, as a consequence of the propagation time difference $\Delta T$, the original and additional comb will be detected with a slightly different value for $f_{\text{rep}}$. A linear scan $f_{\text{rep}}(t)=f_{\text{rep$_0$}}+\alpha \times t$ then leads to $\Delta f(t)=\Delta f=p\alpha\times \Delta T$ constant in time, and thus to a phase difference between the $p^{\th}$ teeth of the two combs after a scan of duration $t$ given by $\Delta\phi(t)=\Delta f \times t$.

The time-separation between original and additional comb pulses, or equivalently the phase delay of the amplitude modulation at $f_{\text{rep}}$ is equal to the phase delay of the $p^{\th}$ teeth (or any other) of the two combs. In the case of a linear scan, it can be expressed as:
\begin{equation}    
\delta_{\text{pulse}}(t)\simeq\dfrac{\Delta \phi(t)}{pf_{\text{rep$_0$}}}=\dfrac{\Delta f \times t}{pf_{\text{rep$_0$}}}=\dfrac{\Delta \nu_{\text{OLO}}(t)}{\nu_{\text{OLO}}}\dfrac{n\Delta L}{c},
\label{Eq_delta_pulse}\end{equation}
\noindent with $\Delta \nu_{\text{OLO}}(t)$ the variation of $\nu_{\text{OLO}}$ over the scan duration t.

\begin{figure}[htp]
	\centering
	\includegraphics[scale=0.22]{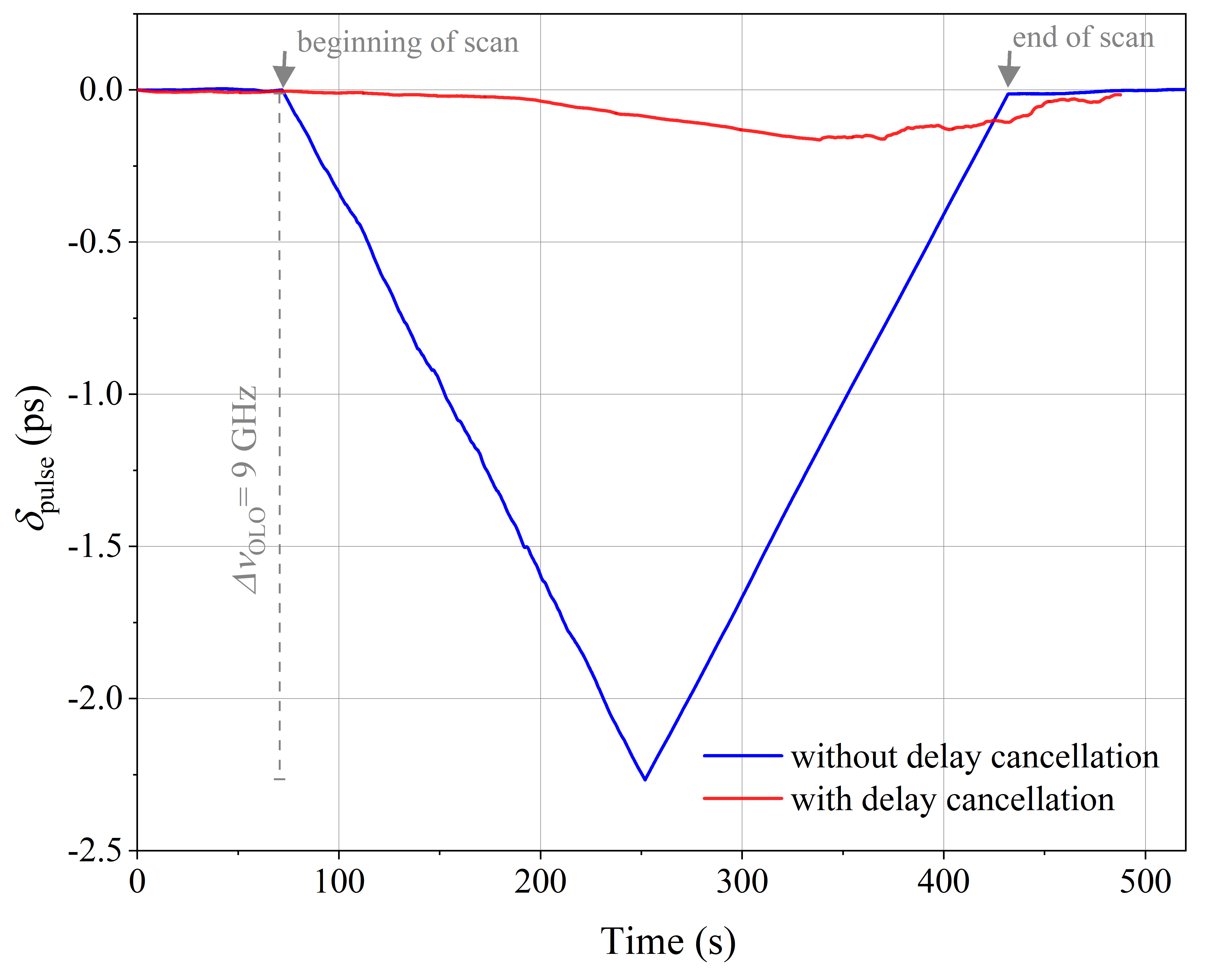}
	\caption{ Evolution of the time-separation between original and additional comb pulses $\delta_{\text{pulse}}(t)$ during a back-and-forth $9$ GHz frequency scan of the OLO frequency: before (blue line : without delay cancellation) and after adding a 10 m-long fiber (red line : with delay cancellation) to the output of the original comb.}
	\label{figA2_2}
\end{figure}

As expected, this corresponds to a a gradual loss of the overlap between comb pulses, corresponding to a time-separation per unit length difference and OLO frequency change of 25~fs/m/GHz. The blue curve in Fig.~\ref{figA2_2} shows $\delta_{\text{pulse}}(t)$ during a back-and-forth 9~GHz OLO frequency scan, before cancellation of the time delay between the original and additional comb. It is determined using Eq.~\ref{Eq_delta_pulse} after numerically integrating $\Delta f(t)$. We observe that $\delta_{\text{pulse}}$ varies by $\sim2.25$~ps during the scan, and conclude that $\Delta L\simeq10$~m.

To compensate the delay between the two combs, we have thus added a 10 m-long fiber to the original comb. We repeat the measurement of $\delta_{\text{pulse}}$ after the adjustment. The result shown in red in Fig.~\ref{figA2_2} confirms a good cancellation of the time delay between the original and additional combs. Residual pulse separations of a few tens of femtoseconds are measured during the scan, which is smaller than the pulses' width. Such fluctuations most probably arise from differential temperature fluctuations between the two paths.

\section{ \label{app_AA} Methanol sample preparation and purity}

We have purchased our methanol from VWR Chemicals ( Méthanol $\geq99.8$\% purity, AnalaR NORMAPUR® ACS, ref. 20847.240). According to the supplier the most abundant impurities in our methanol sample are ethanol and water at the 0.1\% and 0.03\% respectively. Any other impurities are at the 10 ppm level maximum. Methanol is contained in a glass tube connected to the vacuum system. Before each experiment we typically pump on it for about a minute to evacuate any potential gas phase impurities. To our knowledge, no water lines are predicted in the spectral window covered by our spectrum. No other hydrocarbon lines are predicted either, except a few weak to very-weak lines of ethene which do not match with our spectrum. 0.1\% of ethene in our sample which is highly improbable would not even allow us to detect the strongest of those C$_2$H$_4$ lines. After comparing the ethanol and methanol cross-sections provided in the HITRAN database, we conclude that 0.1\% of ethanol would result in ethanol lines 3 orders of magnitude weaker than the strongest observed methanol lines, well below our detection sensitivity. We are thus confident that all the lines probed in this work correspond to methanol rovibrational transitions.

\section{ \label{app_A} Methanol spectroscopic notations}

In this work, we adopt the notations of Ref.~\cite{xu2004new} for the spectroscopic assignment of methanol transitions. The $P(\sigma,\mathrm{co},t,K,J)$ resonances studied here are all $\Delta J = -1$, $\Delta K = 0$ and $\Delta t = 0$ transitions between the ground vibrational state and the first excited state of the C-O stretch vibrational mode $\nu_{8}$, with $\sigma$ , $t$, $K$, and $J$, the torsional symmetry ($A$ or $E$), the torsional mode ($\nu_{12}$) quantum number, and rotational quantum numbers of the lower state ($J$ is the total orbital angular momentum quantum number, $K$ is the quantum number for the projection of the total orbital angular momentum onto the molecular symmetry axis). $K$-doublets of $A$ symmetry have $\pm$ superscript on $K$ to distinguish the $A^{+}$ or $A^{-}$ component of the doublet. For $E$ symmetry levels, $K$ is a signed quantum number with positive (respectively negative) values corresponding to levels often denoted as $E1$ (respectively $E2$).

\section{\label{app_B} Model used for line fitting }

The QCL frequency is modulated at a frequency $f_{\mathrm{m}}$ with a FM amplitude $m$. The instantaneous QCL frequency can be written as $\nu_{\text{QCL}}(t)=\nu_{\mathrm{c}}-m\cos(2\pi f_{\mathrm{m}}t+\Psi)$ with $\nu_{\mathrm{c}}$ the optical carrier frequency and $\Psi$ a phase shift between the frequency modulation and the associated intensity modulation (see below). To achieve a reasonable signal-to-noise ratio, the FM amplitude has been set at $m=250$~kHz, comparable to the measured resonances half-width-at-half-maximum of $\sim380$~kHz resulting in a line shape slightly distorted compared to the typical Lorentzian profile. Furthermore, the residual amplitude modulation associated with FM and the power variation over a scan can both contribute to an asymmetry of the line shape. The power frequency-dependency can result from (i) the QCL gain curve, (ii) the underlying Doppler envelop (potentially non-trivial in case of neighboring lines not resolved in the Doppler regime, see FIG. \ref{fig_2}) or (iii) multi-pass-cell-induced residual interference  fringes not fully averaged out (see text).

We thus fit our data to a model that takes into account both these intensity-modulation-induced asymmetries and the FM-induced distortions. It is based on a profile introduced by Schilt \emph{et al} \cite{schilt2003wavelength} used to fit direct absorption spectra (itself derived from Arndt’s model \cite{arndt1965analytical} which considers only pure FM) that we have straightforwardly adapted to study saturation spectra (by not considering the Beer-Lambert law contribution)\cite{tran2019widely}. The signal detected on the MCT photodetector is a combination of a Gaussian baseline and a saturated absorption signal $s(\nu)$. The latter can be expressed in the following form:
\begin{align}\label{sup1-2}
\begin{split}
s(\nu)&=A\left[B_{1}(\nu_{\mathrm{c}}-\nu_{0})-B_{2}m\cos(2\pi f_{\mathrm{m}}t)+1\right]\\
&\times \frac{1}{\pi}\frac{\gamma}{\left[\nu_{\mathrm{c}}-m\cos(2\pi f_{\mathrm{m}}t+\Psi)-\nu_{0}\right]^{2}+\gamma^{2}},
\end{split}
\end{align} 
which corresponds to a frequency modulated Lorentzian profile multiplied by an amplitude that accounts for the intensity modulation at $f_{\mathrm{m}}$ and the power variation over the scan. $A$ is the line intensity factor, $\nu_{0}$ is the center frequency of the transition, $\gamma$ is the half-width-at-half-maximum line width of the underlying Lorentzian profile, $B_{1}$ and $B_{2}$ are the asymmetry factors related to the quasi-static intensity variations over a scan and to the FM-induced intensity modulation, respectively, and $\Psi$ is the phase shift between intensity modulation and FM. Following the derivation in Ref. \cite{schilt2003wavelength}, the expansion of the signal into a Fourier series gives:
\begin{equation}\label{sup1-3}
s(x)=\frac{A}{\pi\gamma}\left[\sum_{n=0}^{\infty}s_{np}(x)\cos(n2\pi f_{\mathrm{m}}t)-\sum_{n=0}^{\infty}s_{nq}(x)\sin(n2\pi f_{\mathrm{m}}t)\right]
\end{equation}
with $x=(\nu_{\mathrm{c}}-\nu_{0})$ the normalized frequency. The amplitude of the in-phase and quadrature signal (with respect to the intensity modulation) after demodulation on the $n^{th}$ harmonic are respectively given for $n\geq2$ by
\begin{widetext}
\begin{align}\label{sup1-4}
\begin{split}
&s_{np}(x)=(B_{1} x+1)\cos(n\Psi) s_{n}(x)-B_{2}\dfrac{m}{2}\left\{\cos\left[(n-1)\Psi\right]s_{n-1}(x)+\cos\left[(n+1)\Psi\right]s_{n+1}(x)\right\}
\end{split}
\end{align}
\begin{align}\label{sup1-5}
\begin{split}
&s_{nq}(x)=(B_{1} x+1)\sin(n\Psi) s_{n}(x)-B_{2}\dfrac{m}{2}\left\{\sin\left[(n-1)\Psi\right]s_{n-1}(x)+\sin\left[(n+1)\Psi\right]s_{n+1}(x)\right\}
\end{split}
\end{align}
\end{widetext}
Formulas for $n=0$ and 1 are different and given by:
\begin{equation}\label{sup1-4b}
s_{1p}(x)=(B_{1} x+1)\cos\Psi s_{1}(x)-B_{2}\dfrac{m}{2}\left[2s_{0}(x)+\cos(2\Psi)s_{2}(x)\right],
\end{equation}
\begin{equation}\label{sup1-5b}
s_{1q}(x)=(B_{1} x+1)\sin\Psi s_{1}(x)-B_{2}\dfrac{m}{2}\left[\sin(2\Psi)s_{2}(x)\right],
\end{equation}
\begin{equation}\label{sup1-4c}
s_{0p}(x)=(B_{1} x+1)s_{0}(x)-B_{2}\dfrac{m}{2}\cos\Psi s_{1}(x),
\end{equation}
\begin{equation}\label{sup1-5c}
s_{0q}(x)=0.
\end{equation}
\noindent In Eq.~\ref{sup1-4} to \ref{sup1-5c}, valid for any underlying line shape, $s_{n}(x)$ corresponds to the signal obtained after demodulation on the $n^{th}$ harmonic when only considering pure FM (no intensity modulation). In the case of an underlying Lorentzian profile (Arndt's model), it is given by~\cite{arndt1965analytical,schilt2003wavelength}
\begin{equation}\label{sup1-6}
s_{n}(x)=\dfrac{1}{2}(-i)^{n}\epsilon_{n}\gamma\dfrac{\left[\sqrt{(\gamma-ix)^{2}+m^{2}}-(\gamma-ix)\right]^{n}}{m^{n}\sqrt{(\gamma-ix)^{2}+m^{2}}}+c.c.
\end{equation}
where $c.c.$ is the complex conjugate, and $\epsilon_{0}=1$, $\epsilon_{n}=2$ for $n\geq1$. To derive Eq.~\ref{sup1-4} to \ref{sup1-5c}, we have followed the procedure proposed by Schilt \emph{et al.}\cite{schilt2003wavelength}. Our formulas for $s_{np}(x)$ and $s_{nq}(x)$ are equivalent but simpler than those in Schilt \emph{et al.} for $n\geq1$. Their formula for $s_{0p}(x)$ does not agree with our result.

The lock-in amplifier allows us not only to detect the in-phase and quadrature signals but also the signal $s_{n\Phi}(x)$ at any detection phase $\Phi$, with respect to the phase of the intensity modulation
\begin{equation}\label{sup1-7}
s_{n\Phi}(x)=\left[s_{np}(x)\cos\Phi+s_{nq}(x)\sin\Phi\right].
\end{equation}
Our experimental procedure consists in choosing the phase that maximises the signal. At $n^{th}$ harmonic detection, a signal $s_{n\Phi_{n,\mathrm{max}}}(x)$ of maximum amplitude is reached for the detection phase\cite{schilt2003wavelength}
\begin{equation}\label{sup1-8}
\Phi_{n,\mathrm{max}}=n\Psi+k\pi,
\end{equation}
where $k$ is an integer. By introducing Equation \ref{sup1-8} into Equation \ref{sup1-7}, we obtain the signal of maximum amplitude that can be detected for $n\geq2$:
\begin{equation}\label{sup1-9}
s_{n\Phi_{n,\mathrm{max}}}(x)=(B_{1} x+1)s_{n}(x)-B_{2}\dfrac{m}{2}\cos\Psi\left[s_{n-1}(x)+s_{n+1}(x)\right].
\end{equation}
The experimental line shape obtained after demodulation on the first harmonic, as  used in this work is found to be:
\begin{equation}\label{sup1-10}
s_{1\Phi_{1,\mathrm{max}}}(x)=[B_{1} x+1]s_{1}(x)-B_{2}\dfrac{m}{2}\cos\Psi\left[2\; s_{0}(x)+s_{2}(x)\right]
\end{equation}
in which $s_{0}(x)$, $s_{1}(x)$, and $s_{2}(x)$ are given by Equation \ref{sup1-6}.

\section{\label{app_C} Line fitting details }

\begin{figure}
	\includegraphics[width=0.46\textwidth]{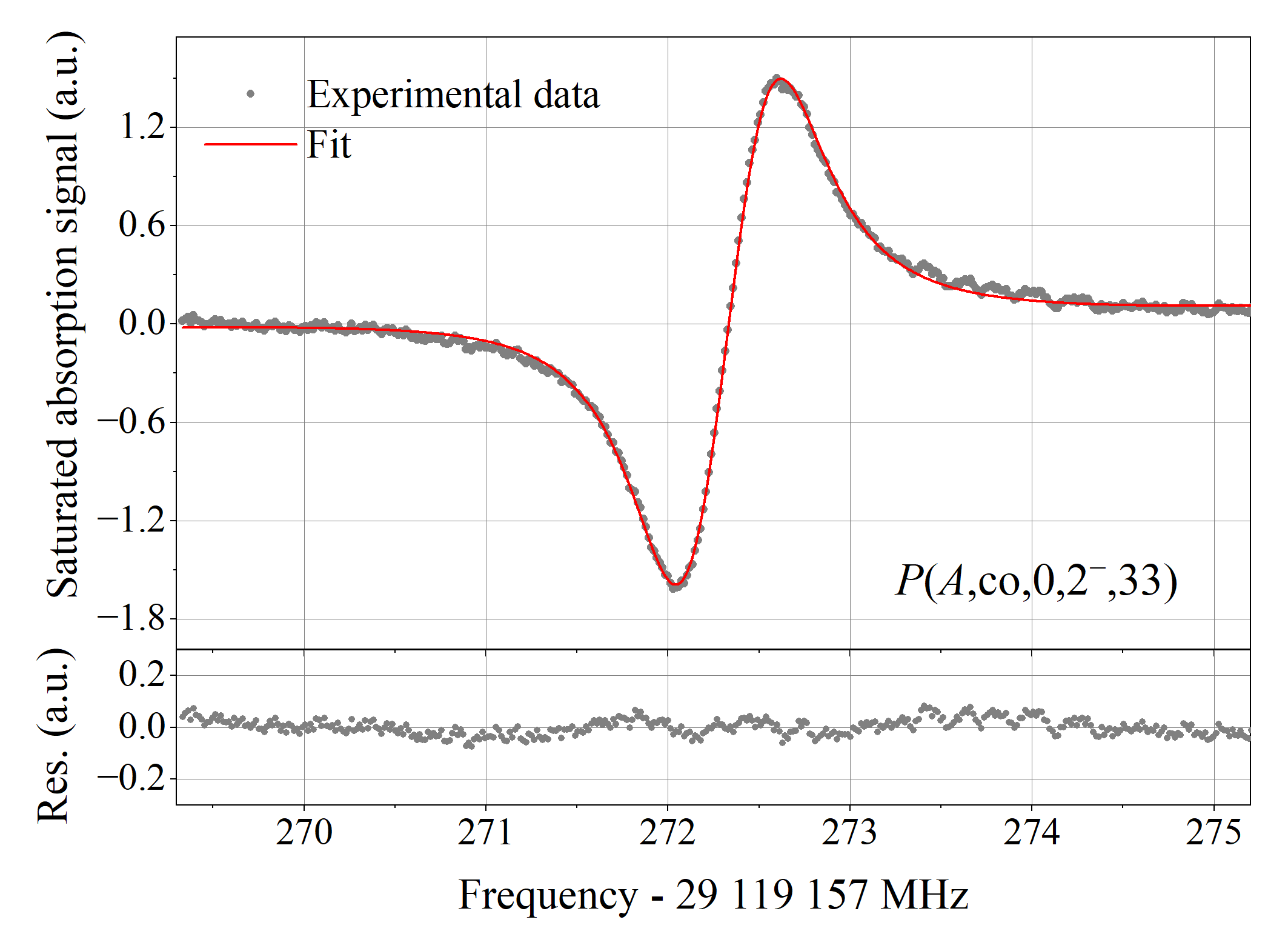}
	\caption{Saturated absorption spectrum of the $P(A,co,0,2^{-},33)$ rovibrational line of methanol, with fit and residuals. The data (grey dots) are recorded using frequency modulation and first-harmonic detection. The red solid line is a fit to the data. Residuals are shown on the bottom graph. Experimental conditions: pressure, 1.5 Pa; modulation frequency, 20 kHz; frequency modulation amplitude, 250 kHz; frequency step, $\sim$15 kHz; average of one pair of up and down scans (corresponding to one of the five pairs contributing to FIG. \ref{fig_2}(b)); total integration time per point, 20 ms; whole spectrum measurement time, 8.02 s.}
	\label{fig_3}
\end{figure}

We fit each averaged pairs of up and down scans of a given transition to the sum of the line profile described in Appendix B and of a second order polynomial to account for the baseline (dominated by the underlying Gaussian contribution on the narrow frequency span used for fitting). All line shape parameters are left free in the fit except the FM amplitude fixed to 250 kHz. The doublets are fitted together using a single second order baseline and, apart from line intensity factors and centre frequencies, line shape parameters set to be equal for the two components. We apply the same procedure for the pair of lines (1) and (2) (see Fig.~\ref{fig_2} (c)). For instance, Fig.~\ref{fig_3} shows the average of one pair of up and down scans of the $P(A,\mathrm{co},0,2^{-},33)$ ro-vibrational transition of methanol (grey points), the fit of the line shape model to the data (red solid curve) and the corresponding residuals. 

\section{\label{app_D} Frequency measurement uncertainty budget }
The reader is referred to Refs. \cite{santagata2019high,tran2019widely} for a detailed description of the line positions uncertainty budget. We only briefly list here the systematic effects and associated uncertainties: (i) we estimate a conservative upper bound on the frequency shift resulting from gas lens effects, wavefront-curvature-induced residual Doppler shifts, the second-order Doppler shift, Zeeman effects, black-body radiation shifts and the photon recoil doublet to be 5 kHz and take this as the corresponding uncertainty; (ii) we assign a conservative 2 kHz systematic uncertainty to the inaccuracy of our model for the line shape (see below); (iii) our 0.02 Pa pressure measurement accuracy and the 5\% specified accuracy of the power meter result in pressure- and power-shift-induced systematic uncertainties of $\sim$0.1 kHz and $\sim$0.5 kHz respectively\cite{santagata2019high} (pressure and power fluctuations respectively smaller than 0.02 Pa and 10\% during a scan and from one to another result in frequency fluctuations of <0.1 kHz and $\sim$1 kHz respectively which are already included in our statistical uncertainty); (iv) the 1 Hz uncertainty on the mid-IR frequency scale is dominated by the $4\times 10^{-14}$ uncertainty on the LNE-SYRTE frequency reference value $\nu_{\text{ref}}$ used to retrieve the absolute frequency scale using Equation (\ref{eq_4}).

We use the three unambiguously assigned highest intensity lines shown in FIG. \ref{fig_2}(a) to determine the systematic uncertainty resulting from the inaccuracy of our model for the line shape (point (ii) above). We fit each averaged pair of up and down scans with the sum of the spectral line shape described in Appendix \ref{app_C} and a polynomial of order $k$ with $k$ ranging from 2 to 5. For all such averaged pairs, we find the standard deviation of the four frequencies resulting from the corresponding fits to be less that 2 kHz, which we take as our conservative systematic uncertainty.


\section*{References}

\bibliography{QCLcomb_Refs}

\end{document}